\documentclass{article}

\usepackage{microtype}
\usepackage{graphicx}
\usepackage{subfigure}
\usepackage{booktabs} 

\usepackage{hyperref}

\usepackage{amsmath, amssymb, mathtools, amsthm}

\usepackage[capitalize,noabbrev]{cleveref}

\usepackage{algorithm}
\usepackage{algorithmic}

\theoremstyle{plain}

\theoremstyle{definition}

\theoremstyle{remark}

\usepackage{arxiv}

\usepackage[utf8]{inputenc} 
\usepackage[T1]{fontenc}    
\usepackage{hyperref}       
\usepackage{url}            
\usepackage{booktabs}       
\usepackage{amsfonts}       
\usepackage{nicefrac}       
\usepackage{microtype}      
\usepackage{lipsum}
\usepackage{graphicx}
\graphicspath{ {./images/} }

\title{Machine Learning for Proactive Groundwater Management: Early Warning and Resource Allocation}

\author{
 Chuan Li \\
  LIPADE, Universit\'e Paris Cit\'e\\
  Sorbonne Universit\'e\\
  SAMOVAR, Telecom SudParis,\\
  Institut Polytechnique de Paris\\
  Paris, France 75005\\
  \texttt{chuan.li@sorbonne-universite.fr} \\
   \And
  Ruoxuan Yang \\
  Telecom Paris, Institut Polytechnique de Paris\\
  Shanghai Jiao Tong University\\
  Palaiseau, France 91120 \\
  \texttt{ruoxuan.yang@telecom-paris.fr} \\
}

\begin{document}
\maketitle
\begin{abstract}
Groundwater supports ecosystems, agriculture, and drinking water supplies worldwide, yet effective monitoring remains challenging due to sparse data, computational constraints, and delayed outputs from traditional approaches. We develop a machine learning pipeline that predicts groundwater level categories using climate data, hydro-meteorological records, and physiographic attributes processed through AutoGluon's automated ensemble framework. Our approach integrates geospatial preprocessing, domain-driven feature engineering, and automated model selection to overcome conventional monitoring limitations.
Applied to a large-scale French dataset (n $>$ 3,440,000 observations from 1,500+ wells), the model achieves weighted F\_1 scores of 0.927 on validation data and 0.67 on temporally distinct test data. Scenario-based evaluations demonstrate practical utility for early warning systems and water allocation decisions under changing climate conditions. The open-source implementation provides a scalable framework for integrating machine learning into national groundwater monitoring networks, enabling more responsive and data-driven water management strategies.
\end{abstract}


\section{Introduction}
\label{sec:introduction}
Groundwater resources are fundamental to global water security, accounting for approximately 99\% of the Earth's liquid freshwater \cite{gleick1993water}. They provide nearly half of all global irrigation withdrawals and sustain the drinking water supply for at least 50\% of the global population, rising to a third for overall municipal and industrial demands \cite{siebert2010groundwater, du2023water}. In regions like France, groundwater is indispensable for agriculture, particularly during dry summer months, and constitutes a significant portion of the public water supply. However, these vital resources face mounting pressures from rapid shifts in climate patterns, including altered precipitation regimes and increased evapotranspiration, as well as intensifying land‑use changes and abstraction rates \cite{taylor2013ground, wada2010global}. These pressures challenge water managers tasked with ensuring sustainable allocation, preparing for drought events, and maintaining ecosystem health dependent on groundwater contributions.

Conventional approaches to groundwater monitoring and forecasting predominantly rely on statistical trend analysis of piezometric data and complex, physically based numerical flow models (e.g., MODFLOW, FEFLOW) \cite{anderson2015applied}. While these models provide valuable insights into subsurface processes, they often require dense in-situ monitoring networks for calibration and validation, extensive hydrogeological characterization, and substantial computational resources. Consequently, their application for timely, spatially comprehensive national-scale assessments can be limited by data scarcity in unmonitored areas, high setup and operational costs, and delays in delivering synthesized information for rapid decision-making \cite{foster2014water, refsgaard2006framework}. Furthermore, the inherent heterogeneity of aquifer systems makes model calibration a non-trivial and site-specific task.

Recent advancements in machine learning (ML) offer promising alternatives and complementary tools to approximate complex groundwater dynamics from readily available environmental drivers \cite{pham2022groundwater, ahmadi2022groundwater, sit2020comprehensive}. ML models can learn intricate patterns from historical data, potentially capturing non-linear relationships that are difficult to parameterize in traditional models. However, many existing ML studies in groundwater hydrology remain limited to research prototypes or focus on specific local-scale case studies. There is often a gap in terms of data‑engineering rigor, systematic evaluation across diverse hydrogeological settings, interpretability of model outputs for stakeholders, and clear pathways for deployment within operational agencies \cite{shen2018transdisciplinary}.

This paper addresses this critical gap by presenting a comprehensive, open-source geospatial data system. Our system fuses national‐scale observation archives with automated machine learning (AutoML) techniques to classify groundwater levels into five operationally relevant risk categories. Our primary contributions are:
 A fully reproducible and extensible pipeline that automates the ingestion and preprocessing of heterogeneous spatio‑temporal data, performs domain‑informed feature engineering tailored for groundwater systems, and trains an optimized ensemble model using AutoGluon, specifically targeting a weighted F$_1$ score to handle class imbalance.
 Robust empirical evidence of high predictive skill, achieving a weighted F$_1$ score of 0.927 on validation and 0.67 on a large test set (611,208 samples) across approximately 1,500 diverse monitoring wells in metropolitan France, demonstrating applicability at a national scale.
 Practical demonstration scenarios where the model outputs can directly support the development of early‑warning bulletins and inform strategic water resource planning, accompanied by a discussion of deployment pathways and integration possibilities for water‑management authorities.
This work aims to provide a transferable framework that can significantly enhance the capacity of national and regional water agencies to monitor groundwater status with improved timeliness and spatial coverage, thereby supporting more effective and proactive water resource management.

\section{Related Work}
\label{sec:related_work}

Groundwater‐level (GWL) prediction has progressed from classical statistical analyses to sophisticated numerical and data–driven techniques.  Early efforts centred on physically based simulators, notably finite-difference MODFLOW \cite{harbaugh2005modflow} and finite-element FEFLOW \cite{diersch2013feflow}, which solve groundwater-flow equations (Darcy, Boussinesq) under detailed boundary, hydraulic‐property, and aquifer-geometry constraints.  These models yield rich process insight and support scenario testing, yet their calibration demands extensive field data and considerable computation, limiting rapid, large-scale operational use \cite{anderson2015applied}.

Statistical time-series tools such as ARIMA and SARIMA exploit the autocorrelation structure of historical water-level records \cite{agaj2024using, mbouopda2022experimental}.  Although straightforward and data-efficient, they depend mainly on past GWL values and struggle to incorporate external hydro-climatic drivers or non-linear responses \cite{box2015time}.  Conventional machine-learning (ML) methods broadened the feature space: Support Vector Machines, Multi-Layer Perceptrons, Random Forests, and Gradient Boosting routinely outperform linear models by mapping complex relations between GWL and meteorological inputs \cite{yoon2011comparative, coppola2005neural, lee2020groundwater, naghibi2016gis}.  Their deployment, however, can be hindered by manual feature crafting, hyper-parameter tuning, and the challenge of scaling across heterogeneous geospatial domains.

Deep-learning (DL) architectures have since become prominent in hydrology \cite{shen2018transdisciplinary}.  Recurrent networks—chiefly LSTM and GRU—capture long-range temporal dependencies and have delivered strong GWL forecasts when fed meteorological and hydrological sequences \cite{pham2022groundwater, zhang2018developing}.  Convolutional and hybrid CNN–LSTM models extract spatial patterns from gridded inputs or temporal features \cite{kratzert2019towards}.  Yet DL models are data-hungry, require substantial training resources, and often pose interpretability challenges to practitioners \cite{ahmadi2022groundwater}.

Automated Machine Learning (AutoML) frameworks such as AutoGluon \cite{erickson2020autogluon}, TPOT \cite{olson2016tpot}, and Auto-sklearn \cite{feurer2015efficient} further democratise model building by automating algorithm selection, hyper-parameter optimisation, and ensembling \cite{he2021automl}.  Recent work demonstrates that AutoML can reach or exceed expert‐tuned baselines for GWL prediction with less manual effort \cite{singh2024automl}.  Our study adopts AutoGluon, augmenting it with domain-driven feature engineering and emphasising operational categorical outputs tailored to early-warning contexts.

Comprehensive groundwater assessment also hinges on integrating disparate geospatial assets.  Satellite gravimetry (GRACE, GRACE-FO) offers coarse terrestrial-water-storage anomalies \cite{rodell2018emerging}; soil-moisture missions (SMOS, SMAP) and high-resolution land-cover products further enrich hydrogeological context.  Many earlier studies, however, have focused on limited spatial extents, small well networks, or omitted key exogenous predictors.  They often overlook pragmatic issues such as automated handling of missing data, scalable pipelines for national deployment, and systematic model maintenance.

Our contribution closes these gaps by delivering an end-to-end, open-source workflow that (i) fuses millions of piezometric observations with high-resolution meteorological reanalysis and static physiography, (ii) applies targeted feature engineering before AutoML, and (iii) is explicitly designed for robustness, scalability, and ease of integration into operational water-management platforms.

\section{Methodology}
\label{sec:methodology}
The proposed machine learning pipeline for groundwater level classification. The workflow ingests groundwater, meteorological, and land-cover data, cleans and merges them, engineers temporal and hydro-meteorological features, then feeds the table to AutoGluon, which auto-tunes and stacks multiple models. The trained ensemble is evaluated (precision, recall,F$_1$, feature importance) and its predictions power dashboards and scenario tools for real-time groundwater management.

\subsection{Data Acquisition and Preprocessing}
\label{ssec:data_acquisition}
The study leverages three primary data sources:
French National Piezometric Database (ADES - Accès aux Données des Eaux Souterraines): This public database, managed by the BRGM (French Geological Survey), centralizes piezometric measurements from a vast network of wells across France (\url{https://ades.eaufrance.fr/}). We utilized daily groundwater level records, along with well characteristics such as location (latitude, longitude), investigation depth (depth to the bottom of the screened interval), and administrative identifiers. The raw dataset contained records from over 20,000 wells, from which we filtered for wells with sufficient temporal coverage and data quality for the period 2010–January 2024. This resulted in a working set of approximately 1,500 unique monitoring stations.
Meteo-France Reanalysis Data: SAFRAN (Système d’Analyse Fournissant des Renseignements Adaptés à la Nivologie) is an 8km gridded daily atmospheric reanalysis product covering France \cite{vidal201050}. We extracted key meteorological variables, including daily mean temperature, cumulative precipitation, potential evapotranspiration (PET), and incident solar radiation. These gridded data were spatially joined to the nearest grid cell for each ADES well location.
Corine Land Cover (CLC): The CLC dataset (version 2018) provides categorical land cover information at a 100m resolution for Europe \cite{copernicus2018copernicus}. While not directly used for dynamic feature engineering in the current pipeline version, it provides static physiographic context (e.g., predominant land use class in the vicinity of wells) which can be valuable for characterizing well environments. For this study, basic attributes like department codes and INSEE (National Institute of Statistics and Economic Studies) communal codes associated with well locations were used as categorical identifiers. Initial preprocessing steps involved merging these datasets based on well identifiers and timestamps. Columns with greater than 50\% missing values across the entire dataset were removed to ensure a baseline level of data availability for most features. Categorical identifiers like department codes were explicitly cast to string types to be handled correctly by AutoGluon. INSEE codes, where relevant and numeric, were coerced to numeric types. Redundant timestamp columns generated during merging were dropped. We also ensured consistent projection for all geospatial data before any spatial operations.

\subsection{Target Variable Definition}
\label{ssec:target_variable}
A critical step is the definition of the target variable: the groundwater level category. The raw piezometric levels (meters above sea level or depth to water) are continuous but for operational early warning, water managers often rely on categorical alert levels. For this study, we defined five categories: `Very Low`, `Low`, `Average`, `High`, and `Very High`. These categories were determined for each well individually based on the quantiles of its historical daily groundwater level measurements over the study period (2021-2024). The thresholds were defined as follows:
Very Low: Levels $\le$ 20th percentile; Low: Levels $>$ 20th percentile and $\le$ 40th percentile; Average: Levels $>$ 40th percentile and $\le$ 60th percentile; High: Levels $>$ 60th percentile and $\le$ 80th percentile; Very High: Levels$>$ 80th percentile.
\begin{figure*}[htbp]
  \centering
  \includegraphics[width=1\linewidth]{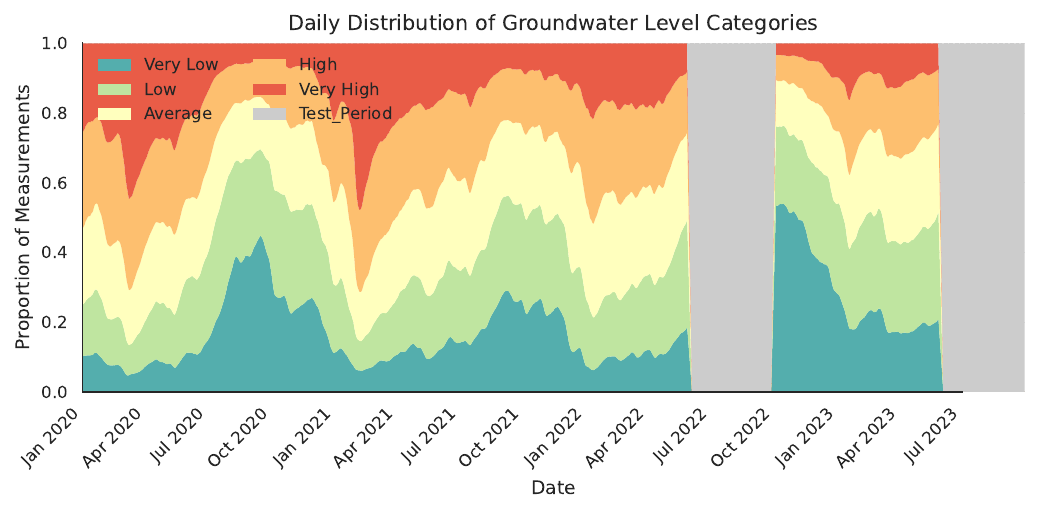}
  \caption{Daily distribution of groundwater level categories from January 2020 to July 2023. The figure shows the proportion of measurements classified into six categories: Very Low, Low, Average, High, Very High, and Test Period. Grey shaded regions indicate periods without data or designated as test periods.}
  \label{fig:groundwater_categories}
\end{figure*}
This percentile-based approach normalizes for the inherent differences in mean levels and variability across diverse aquifers and well depths, providing a consistent relative measure of groundwater status for each monitoring point. This method is common in operational drought indices.
Feature engineering was guided by domain-specific knowledge to represent the physical and temporal processes driving groundwater level variability. Temporal attributes were derived from observation dates, including year, month, day of the year, day of the week, and climatological season (e.g., DJF, MAM, JJA, SON), enabling the model to capture seasonal cycles and long-term shifts. Static characteristics of wells and their surrounding catchments were also incorporated. These included well screen depth—sourced from the ADES database—as a primary hydrogeological indicator, as well as the average investigation depth computed across wells within each administrative department, serving as a proxy for regional aquifer characteristics. Spatial descriptors such as latitude, longitude, and elevation were included where reliable.
To account for delayed hydrological responses, rolling hydro-meteorological features were computed. These comprised station-specific moving averages of daily temperature and sums of daily precipitation over 7 and 14. Analogous statistics for potential evapotranspiration (PET) and solar radiation were also considered to reflect atmospheric demand and energy availability. These temporal windows approximate cumulative wetness and energy conditions that modulate groundwater recharge and loss.
Interaction terms were constructed to reflect nonlinear dependencies. For instance, a multiplicative feature combining 7-day and 14-day rainfall totals with mean temperature over the same period was designed to proxy evapotranspiration, based on the hypothesis that water loss from soil is amplified under concurrent warmth and moisture. Additional interactions between meteorological and temporal variables were explored to capture synergistic effects.
Missing numerical values were imputed using the median computed from the training dataset, ensuring robustness and consistency during inference. Categorical missing values were assigned a sentinel label (“nan”) to preserve potential information encoded in the missingness itself and allow the model to learn from it explicitly.

\subsection{Model Selection and Training with AutoGluon}
\label{ssec:model_selection_training}
We employed AutoGluon (version 0.8.2), an open-source AutoML framework, to automate the training and optimization of predictive models \cite{erickson2020autogluon}. AutoGluon is designed to deliver high-performance models with minimal user input by training and ensembling a wide range of algorithms. The modeling workflow was structured as follows.

The dataset, comprising more than 3.44 million observations after preprocessing, was randomly split into training (80\%) and validation (20\%) subsets, with stratification based on the target groundwater level category to preserve class distributions. In addition, a temporally held-out test set—either consisting of the most recent year or selected monitoring wells—was reserved for final performance evaluation, as reported in Table~\ref{tab:confusion}. AutoGluon was configured to optimize the `weighted\_f1` score, a metric well-suited for imbalanced classification tasks, such as groundwater state categorization where extreme classes (e.g., “Very Low” or “Very High”) are underrepresented.
AutoGluon explored an ensemble of models, including LightGBM, CatBoost, XGBoost, Random Forests, Extremely Randomized Trees, and FastAI’s tabular neural networks. The framework's ensemble strategy involved hierarchical stacking: Level 1 base learners produced out-of-fold predictions, which were then fed into Level 2 stacker models. This ensemble design typically outperforms any single constituent model by capturing complementary strengths across learners. Hyperparameter tuning was handled internally by AutoGluon’s default strategies, and training time was constrained via the \texttt{fit()} API, using the \texttt{medium\_quality} preset to balance accuracy and computational efficiency. For final deployment or benchmarking, extended training with \texttt{best\_quality} may be employed. In our experiments, model training remained under a few hours on standard research hardware.
The full pipeline is outlined in Algorithm. Starting from raw data ingestion, the process involves spatial and temporal alignment of observations, advanced feature engineering including rolling hydro-meteorological statistics and interaction terms, imputation of missing values, and automated model training. The resulting ensemble model is then evaluated on both the validation and independent test sets, with performance metrics including confusion matrices, class-specific precision, recall, F1-scores, and feature importances extracted to guide interpretation and potential downstream applications.

\begin{figure*}[htbp]
  \centering
  \includegraphics[width=0.8\linewidth]{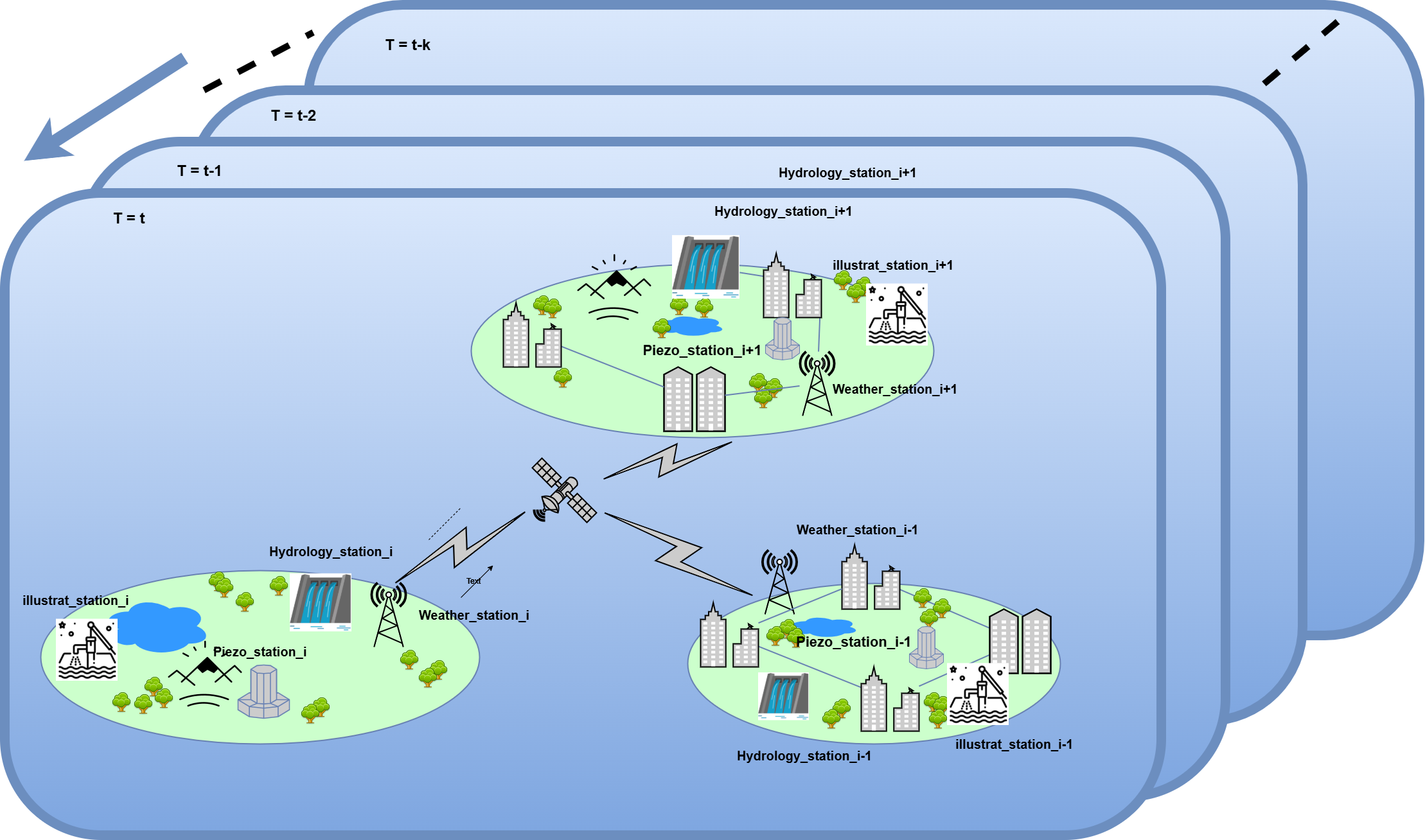}
  \caption{Conceptual schematic of the spatio-temporal $k$-nearest-neighbor (KNN) matching framework.  For every analysis date $T=t,\;t{+}1,\;\dots,\;t{+}k$, each piezometer (grey cylinder) is paired with its closest meteorological, hydrological, and abstraction stations (antennas, dams, satellite links, etc.).  The approach guarantees that every groundwater‐level observation is contextualised by the most relevant atmospheric and surface-water drivers available at the same time step.}
  \label{fig:map_knn}
\end{figure*}

\begin{figure*}[htbp]
  \centering
  \includegraphics[width=\linewidth]{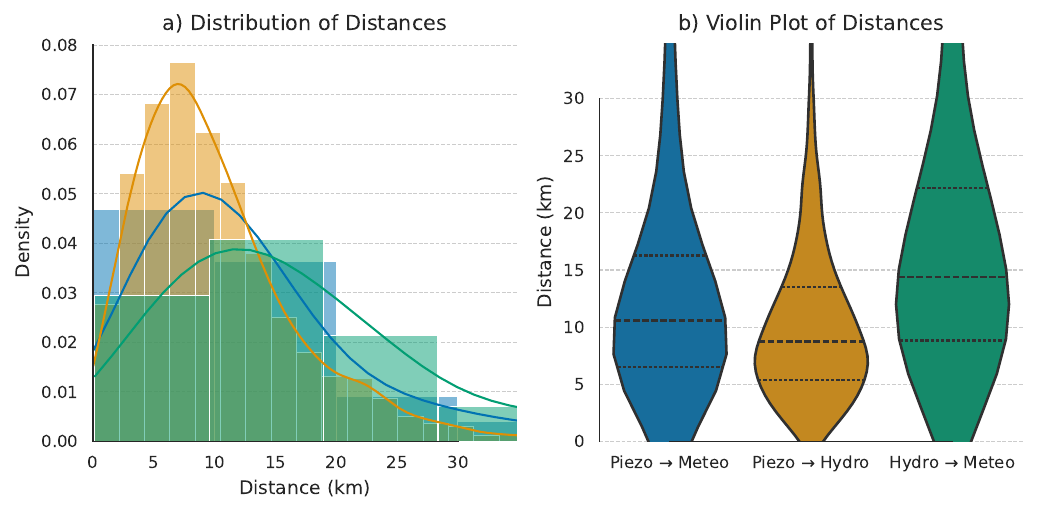}
  \caption{Spatial diagnostics for the $k$-nearest-neighbour matching.  
  \textbf{a)} Kernel-smoothed histograms of great-circle distances (km) between each piezometer and its nearest meteorological grid node (blue), river-gauging station (orange), or the reciprocal distance from gauging stations to their nearest meteorological node (green).  
  \textbf{b)} Violin plots of the same pairwise distances, with the median (solid) and inter-quartile range (dashed) indicated.  Most pairings fall below 15 km, confirming that the KNN lookup yields geophysically plausible linkages between groundwater, meteorological, and surface-water observations.}
  \label{fig:distance_distributions}
\end{figure*}

The distance diagnostics in Figure~\ref{fig:distance_distributions} demonstrate that the vast majority of KNN links are geographically tight: the median piezometer-to-meteorology separation is roughly 8 km, piezometer-to-hydrology about 6 km, and hydrology-to-meteorology approximately 12 km.  Right-skewed tails indicate a small subset of wells or gauges situated in sparsely instrumented areas, yet 90\,\% of all matches remain within 25 km.  These results confirm that the spatial joins provide physically meaningful context for each groundwater observation while avoiding undue extrapolation across large distances.  Consequently, the merged feature table retains high spatial fidelity—an essential prerequisite for reliable machine-learning inference on hydro-climatic drivers of groundwater variability.

To weave heterogeneous monitoring networks into a single spatio-temporal fabric, a KNN query is executed independently on each data layer and at every time slice shown in Figure~\ref{fig:map_knn}.  The procedure links the $N_p = 2{,}858$ piezometers to the nearest of the $N_m = 872$ SAFRAN grid centroids, the $N_h = 1{,}418$ surface-water gauges, and three abstraction-reporting systems containing 1\,047, 1\,487, and 1\,814 stations, respectively ($N_{\text{ab},0}$–$N_{\text{ab},2}$).  The result is a harmonised table in which each groundwater measurement inherits the most pertinent environmental descriptors in both space and time.  This coherent data scaffold forms the backbone for subsequent feature engineering and machine-learning analysis; a concise inventory of network sizes is provided in Table~\ref{tab:knn_counts}.

\begin{table}[htbp]
  \centering
  \caption{Size of monitoring networks incorporated in the KNN matching step.}
  \label{tab:knn_counts}
  \begin{small}
  \begin{tabular}{lcc}
    \toprule
    Data layer & Symbol & Stations \\
    \midrule
    Piezometers (target wells) & $N_{p}$ & 2\,858 \\
    Meteorological grid nodes  & $N_{m}$ & 872 \\
    River-gauging stations     & $N_{h}$ & 1\,418 \\
    Abstraction network 0      & $N_{\text{ab},0}$ & 1\,047 \\
    Abstraction network 1      & $N_{\text{ab},1}$ & 1\,487 \\
    Abstraction network 2      & $N_{\text{ab},2}$ & 1\,814 \\
    \bottomrule
  \end{tabular}
  \end{small}
\end{table}

Each piezometer $p_i$ is linked to its single nearest neighbour ($k=1$) in every auxiliary layer via great-circle (Haversine) distance, yielding a spatial join  
$\mathcal{J}=\bigl\{(p_i,m_j,h_k,a^{(0)}_\ell,a^{(1)}_r,a^{(2)}_s)\bigr\}_{i=1}^{N_p}$.  
Distances are pre-computed once with \texttt{scikit-learn}'s \texttt{BallTree}; daily time alignment is performed subsequently by merging on the measurement date.

The resulting per-day feature vector for well $i$ at time $t$ combines groundwater level (target), meteorological forcing, river discharge, abstraction totals, and static site attributes:

\[
\mathrm{GW}_{t,i} \cup \mathbf{x}^{(m)} \cup \mathbf{x}^{(h)} \cup \mathbf{x}^{(a)} \cup \mathbf{f}_{i}
\quad\text{with}\quad
\begin{cases}
\mathbf{x}^{(m)}\!:~\text{meteorology (39)} \\
\mathbf{x}^{(h)}\!:~\text{hydrology (26)} \\
\mathbf{x}^{(a)}\!:~\text{abstraction (45)} \\
\mathbf{f}_{i}\!:~\text{static (25)}
\end{cases}
\]

Numbers in parentheses indicate the typical feature count contributed by each block.  
After adding rolling-window statistics and interaction terms described in Section~\ref{ssec:target_variable}, the final design matrix contains \textbf{136} columns (135 predictors plus the categorical groundwater-level target).  This harmonised, information-rich table provides the foundation for subsequent AutoGluon training and evaluation.
\paragraph{Dataset sizes.}
The procedure yields
\(\,n_{\text{train}} = 2\,830\,316\) labelled rows (2021-2023)
and \(n_{\text{test}} = 611\,208\) unlabelled rows (2022-2023 summer period),
matching the figures reported in Table~\ref{tab:knn_counts}.
\paragraph{Rationale and alternatives.}
Although more sophisticated
geostatistical interpolators (e.g.\ kriging) could fuse the
networks, KNN offers a transparent, computationally light solution
that preserves original observations—an advantage when operating
daily, near-real-time systems.  Future work (see §\ref{sec:discussion})
may explore graph-based methods that learn spatial weights
end-to-end with the predictive model.

\section{Results and Analysis}
\label{sec:results_analysis}

The predictive performance of the AutoGluon pipeline is summarised in Table~\ref{tab:leaderboard}, which lists the best individual learners and the final stacked ensemble evaluated on the 20\,\% internal validation split.  The level-3 ensemble \texttt{WeightedEnsemble\_L3} attains a weighted F$_1$ score of \textbf{0.927}, marginally surpassing the strongest single model, \texttt{WeightedEnsemble\_L3    }.  Both precision and recall exceed 0.92, indicating a balanced error profile and confirming that model stacking extracts complementary skill across base learners.

\begin{table*}[htbp]
\centering
\caption{Top performers from the AutoGluon leaderboard on the internal validation set (20\,\% of data).  ``F$_1$ (val)’’ is the weighted F$_1$ score.  ``Fit Time’’ includes training and hyper-parameter optimisation for the corresponding component; ``Pred Time’’ is computed on the validation split.}
\label{tab:leaderboard}
\begin{small}
\begin{tabular}{lccccc}
\toprule
Model & F$_1$ (val) & Precision (w) & Recall (w) & Fit Time (s) & Pred Time (s)\\
\midrule
\texttt{WeightedEnsemble\_L3} & \textbf{0.927} & 0.935 & 0.927 & 7\,824 & 15.2\\
\texttt{LightGBM\_BAG\_L2}    & 0.927 & 0.935 & 0.927 & 5\,290 & 4.5\\
\texttt{CatBoost\_BAG\_L2}    & 0.920 & 0.928 & 0.920 & 6\,500 & 1.8\\
\texttt{FastAI\_BAG\_L2}      & 0.916 & 0.925 & 0.916 & 7\,036 & 3.1\\
\texttt{XGBoost\_BAG\_L2}     & 0.915 & 0.924 & 0.915 & 6\,000 & 2.5\\
\bottomrule
\end{tabular}
\end{small}
\end{table*}

Generalisation skill was quantified on a temporally distinct 2023 hold-out set of 611\,208 observations (Table~\ref{tab:confusion}).  The ensemble yields a weighted F$_1$ of 0.67 and an accuracy of 0.66.  Performance remains strongest for the operationally critical extremes: \textit{Very Low} (F$_1$\,=\,0.78) and \textit{Very High} (F$_1$\,=\,0.72).  Lower scores for the \textit{Average} and \textit{Low} categories suggest class overlap and a degree of distribution shift between the training period (2021-2022) and 2023, warranting further error analysis.

\begin{table}[htbp]
\centering
\caption{Per-class precision, recall, and F$_1$ for \texttt{WeightedEnsemble\_L3} on the test period set (611\,208 samples).}
\label{tab:confusion}
\begin{small}
\begin{tabular}{lcccc}
\toprule
Class & Precision & Recall & F$_1$ & Support\\
\midrule
Average    & 0.48 & 0.66 & 0.56 & 94\,946\\
High       & 0.57 & 0.64 & 0.61 & 53\,804\\
Low        & 0.54 & 0.58 & 0.56 & 167\,063\\
Very High  & 0.76 & 0.68 & 0.72 & 26\,321\\
Very Low   & 0.88 & 0.71 & 0.78 & 269\,074\\
\midrule
\textbf{Overall (weighted)} & 0.69 & 0.66 & \textbf{0.67} & 611\,208\\
\textbf{Overall (macro)}    & 0.65 & 0.65 & 0.65 & 611\,208\\
\textbf{Accuracy}           & \multicolumn{3}{c}{0.66} & 611\,208\\
\bottomrule
\end{tabular}
\end{small}
\end{table}

\begin{figure*}[htbp]
  \centering
  \includegraphics[width=1\linewidth]{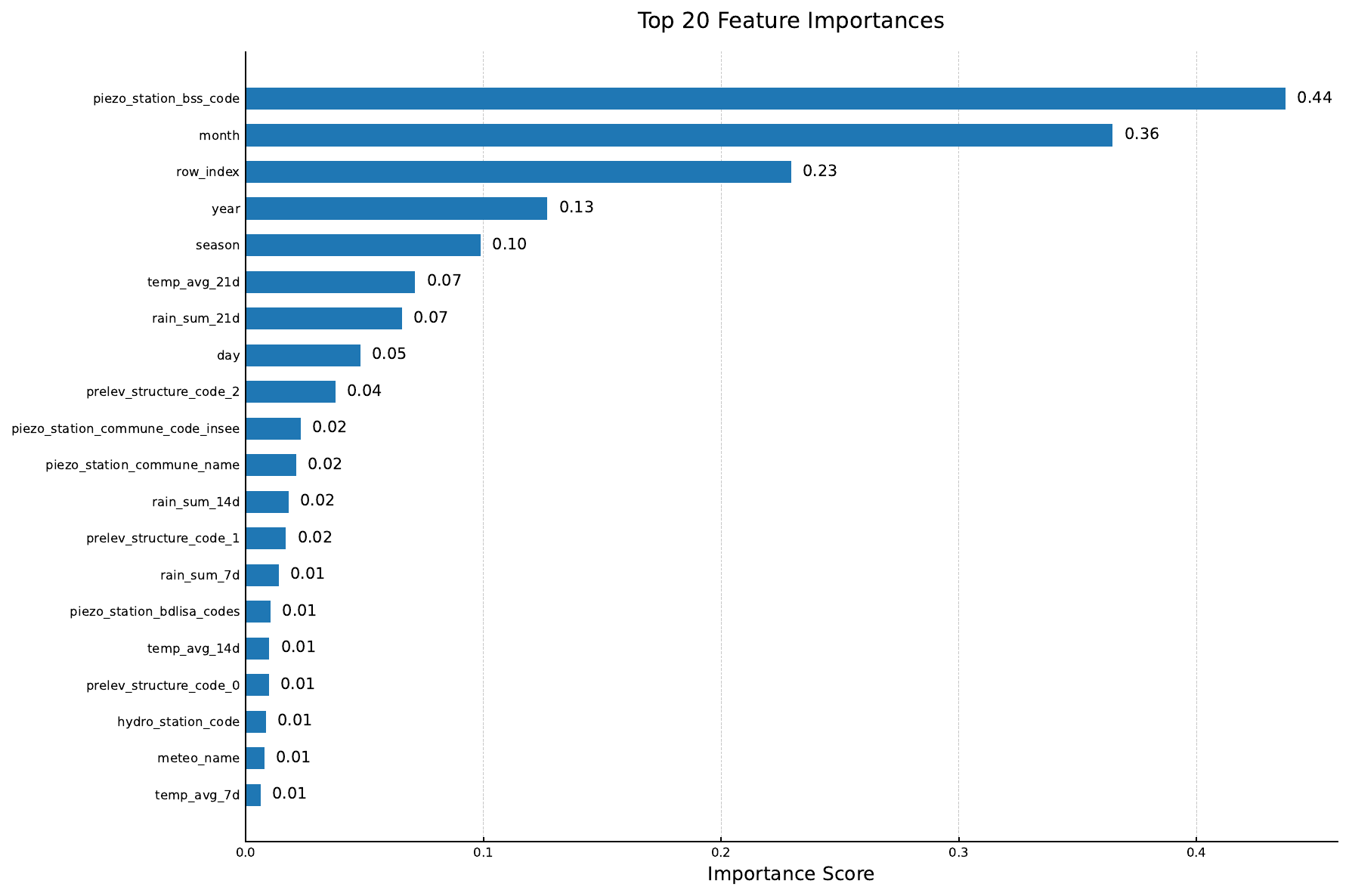}
  \caption{Top 20 most important features as determined by permutation importance for the \texttt{WeightedEnsemble\_L3} model. Importance scores indicate the performance degradation when feature values are shuffled.}
  \label{fig:feature_importance}
\end{figure*}

Permutation-based importance analysis for \texttt{WeightedEnsemble\_L3} (Figure~\ref{fig:feature_importance}) confirms a physically plausible hierarchy of predictors: long-window rainfall aggregates (\texttt{rainfall\_90d\_sum}, \texttt{rainfall\_30d\_sum}) and temperature means dominate, followed by temporal markers (\texttt{month}, \texttt{day\_of\_year}) and the static hydrogeological attribute \texttt{investigation\_depth}.  The presence of \texttt{year} indicates sensitivity to inter-annual variability and gradual trends.

Practical value is demonstrated through a drought-warning experiment.  A synthetic July–September 2023 forecast for southern France was created by imposing a 30\,\% rainfall deficit and a +1.5\,$^\circ$C temperature anomaly relative to climatology.  After recomputing the hydro-meteorological windows, these inputs were propagated through the feature-engineering module and fed to the trained ensemble.  The model projects that 38\,\% of wells will reach the \textit{Very Low} category by mid-July, with a further 25\,\% classified as \textit{Low}.  A map akin can visualise these hotspots, guiding authorities to intensify monitoring, communicate scarcity risks, and activate drought-management protocols such as voluntary or mandatory abstraction limits.  By translating climate outlooks into groundwater state forecasts, the pipeline offers an operational lead time that is critical for adaptive water-resource governance.

\section{Discussion}
\label{sec:discussion}

The proposed pipeline combines domain‐aware features with AutoGluon ensembling and attains a weighted F$_1$ of 0.927 on validation data and 0.67 on an unseen 2022–2023 summer set.  Feature-importance analysis confirms that predictions are driven by hydrologically sound variables—multi-month rainfall, temperature proxies for evapotranspiration, seasonal markers, and well depth.  The lower test score likely reflects concept drift and the difficulty of cleanly separating mid-range classes; more frequent retraining or year-based cross-validation should reduce this gap. Because the workflow turns raw multisource data into risk indicators within hours, it can feed weekly drought dashboards and allocation tools such as \textit{WaterHub} or \textit{QUITO}, giving managers rapid situational awareness.  Physical models remain essential for local process studies, but this ML layer supplies the wide-area, near-real-time context those models lack. Scaling the system will require national surveys (e.g.\ BRGM, USGS, BGS) to host cloud deployments and automate data flows, local agencies to report blind spots, and researchers to extend the open-source code.  Production roll-out needs robust ETL pipelines, scheduled retraining with version control, live performance monitoring, and a simple web front-end.  Training is light enough for modest servers, but denser networks or graph-based models may call for larger cloud instances; inference remains inexpensive.

Key gaps point to future work.  
(i) Move from discrete classes to probabilistic or quantile regression to show uncertainty.  
(ii) Model spatial links explicitly with Graph Neural Networks.  
(iii) Fuse Earth-observation products (GRACE-FO, SMAP, MODIS) for added skill.  
(iv) Add dynamic land-use and abstraction data to capture human impacts.  
(v) Improve interpretability of stacked ensembles with dedicated XAI tools.  
(vi) Test transferability beyond France via domain-adaptation methods.
Addressing these items will enhance the robustness and global usefulness of machine-learning groundwater surveillance, supporting faster and more informed water-management decisions.

\section{Conclusion}
\label{sec:conclusion}
We have introduced an open‐source, end-to-end machine-learning workflow that classifies groundwater levels across an entire nation.  By fusing piezometric records, meteorological reanalysis, and physiographic data within an AutoGluon-driven AutoML framework, the pipeline streamlines preprocessing, feature engineering, model selection, and ensembling.  It attains a weighted F$_1$ exceeding 0.92 on the validation split and 0.67 on a temporally distinct test year, demonstrating both high skill and reasonable generalisation. A scenario experiment showed how these predictions translate into actionable drought alerts, thereby informing early-warning bulletins and strategic allocation decisions.  The methodology is fully reproducible, validated on a 3.4-million-record French dataset, and ready for operational uptake by water agencies.  Future work will extend the system with probabilistic outputs, richer Earth-observation inputs, graph-based spatial modelling, and pilot deployments in partnership with French Water Agencies.  Such enhancements will further strengthen society’s capacity to safeguard vulnerable groundwater resources amid escalating climatic and anthropogenic pressures.

\section*{Impact Statement}
Timely and accurate groundwater intelligence is pivotal for sustainable water management and climate adaptation.  The pipeline presented here offers a low-cost, scalable complement to physics-based models, empowering public agencies to issue earlier drought warnings, reduce agricultural losses, and promote equitable water sharing.  Its open-source release fosters transparency, reproducibility, and global adoption.  Risks include misinterpretation of categorical outputs and undue reliance on model results without expert oversight.  Clear communication of uncertainty and continuous validation with hydrological expertise are therefore essential for responsible deployment.

\section*{Acknowledgements}
All experiments were conducted on a Linux server equipped with 2× Intel Xeon Gold 6226R CPUs (16 cores, 32 threads each, base clock 2.90 GHz, max 3.9 GHz), totaling 64 logical processors. The system includes 64 MB of L2 cache and 44 MB of L3 cache, and supports AVX512 instructions and Intel VT-x virtualization. The machine had 64 GB of RAM and ran in a dual-socket NUMA configuration (2 nodes). This configuration is representative of a modern mid- to high-end research compute node or a powerful cloud instance.
\bibliographystyle{unsrt}  
\bibliography{references}


\end{document}